# The causal meaning of genomic predictors and how it affects the construction and comparison of genome-enabled selection models

Bruno D. Valente [1]*§, Gota Morota §, Guilherme J.M. Rosa §†,

Daniel Gianola*§†, Kent Weigel*

* Departments of Dairy Science; § Animal Sciences, † Biostatistics and Medical Informatics, University of Wisconsin-Madison, Madison, WI, 53706.

---

[1] Animal Sciences Building, 1675 Observatory Dr., Madison - WI 53706.

# ABSTRACT

The additive genetic effect is arguably the most important quantity inferred in animal and plant breeding analyses. The term *effect* indicates that it represents causal information, which is different from standard statistical concepts as *regression coefficient* and *association*. The process of inferring causal information is also different from standard statistical learning, as the former requires causal (i.e. non-statistical) assumptions and involves extra complexities. Remarkably, the task of inferring genetic effects is largely seen as a standard regression/prediction problem, contradicting its label. This widely accepted analysis approach is by itself insufficient for causal learning, suggesting that causality is not the point for selection. Given this incongruence, it is important to verify if genomic predictors need to represent causal effects to be relevant for selection decisions, especially because applying regression studies to answer causal questions may lead to wrong conclusions. The answer to this question defines if genomic selection models should be constructed aiming maximum genomic predictive ability or aiming identifiability of genetic causal effects. Here, we demonstrate that selection relies on a causal effect from genotype to phenotype, and that genomic predictors are only useful for selection if they distinguish such effect from other sources of association. Conversely, genomic predictors capturing non-causal signals provide information that is less relevant for selection regardless of the resulting predictive ability. Focusing on covariate choice decision, simulated examples are used to show that predictive ability, which is the criterion normally used to compare models, may not indicate the quality of genomic predictors for selection. Additionally, we propose using alternative criteria to construct models aiming for the identification of the genetic causal effects.

**Introduction**

Many quantities inferred in animal and plant breeding studies are typically labeled as "effects". One of them is the (arguably) most important information inferred in these areas: the additive genetic effect. This information is also known as breeding value (both terms will be applied interchangeably throughout this manuscript). The term "effect" in "genetic effect" suggests a causal meaning. Nevertheless, the inference of genetic effects is normally presented as a statistical problem. It should be stressed, though, that there are important differences between statistical concepts like "association" and "regression coefficient" and causal concepts such as "effect" (Pearl 2000; Pearl 2003). Statistical concepts are limited to describing aspects of a joint probability distribution, from which they can be completely deduced. Conversely, the description of an effect of a variable $x$ on another variable $y$ is related to the change in the expected value of $y$ if the value of $x$ was physically modified. This concerns not only how variables are statistically related, but also how they are structurally or mechanistically related. Probability distributions do not carry this sort of information, following necessarily that the causal relationship among variables cannot be deduced from their joint distribution alone. Therefore, associations and effects have different meanings and require different learning processes.

Causal relationships among traits may be seen as more informative than statistical relationships. However, the objectives of many studies may not require learning causal effects. In biology, medicine and agriculture, joint distributions may be explored for the construction of regression models with the sole goal of predicting future realizations of one variable conditionally on the values observed for other variables. In this case, distributions carry all the relevant information for the task. For example, if one is interested in predicting a trait related to reproductive efficiency conditionally on the blood levels of a specific hormone, then a non-null association between them already indicates that the prediction is possible. In this case, a joint distribution is sufficient information to build a predictor (e.g. by deriving conditional expectations). However, goals of investigations frequently involve learning causal rather than statistical information. If the aspiration was to learn *if* and *how much* reproductive efficiency can be improved by the inoculation of a specific hormone (i.e. by an external intervention on blood hormone levels), then this information could not be deduced from a joint distribution alone even if these variables were highly associated. One reason for the insufficiency is that different causal relationships between hormone levels and reproductive performance could result in the same observed association, although each of these hypothetical causal relationships would imply a different response to the described intervention (Pearl 2000; Spirtes *et al.* 2000; Rosa and Valente 2013).

While randomized trials (Fisher 1971) are benchmark for causal effect inferences, learning about causality from observational data is more challenging. Such task is possible, but it is more complex than standard statistical inference. In addition to all the aspects that are also pertinent to statistical learning, it requires considering non-statistical assumptions and features on how variables are related. Furthermore, this additional information involves concepts that are generally less straightforward than statistical concepts (e.g. intervention, confounding, influence and so forth). It should be stressed, however, that many times a randomized study is not possible. Moreover, a standard regression study is not an alternative when inferring an effect is the goal, regardless of how easier it is to formulate and accept its assumptions, since it provides information that does not satisfy the objective.

Although the terms "effect" and "regression coefficient" present different meanings and require different assumptions for inference, they are many times perceived as similar and used interchangeably. As a result, one might unwittingly (and unbecomingly) adopt the approach to infer one type of information when the goal is to infer the other type. For example, one may indeed need to infer a causal effect to fulfill the study objectives (i.e. the regression approach is not sufficient) but may treat the investigation as a regression (prediction) problem, ignoring the required assumptions for causal inferences. This is a dangerous conduct since conclusions obtained from projecting a causal meaning to results may be wrong if causal assumptions that support them do not hold.

One (at least apparent) incongruence in animal and plant breeding analyses is that in one hand the term "effect" in "genetic effect" suggests that the inference of this important quantity pertain to the realm of causal inference from observational data. On the other hand discussions regarding the requirements of this type of inference, or even the terminology required to articulate this discussion are virtually absent in studies on these fields, suggesting that causality is not the point. Given this conflict and the risk involved in managing an analysis for causal inference as a regression analysis (and vice-versa), it is important to resolve the following issue: could genomic predictors be treated as standard predictors from a regression model, or should they be treated necessarily as reflecting causal effects to be useful? If the regression approach is sufficient, we do not need to worry about the difficulties inherent to causal inference. On the other hand, if they do need to be treated as causal effects, then a large set of new questions arise: Do the predictors obtained from any mixed model reflect this causal effect, or is confounding possible? What are the causal assumptions to identify genetic effects from these predictors? How much of the identifiability of genetic causal effects from predictors is evaluated from the criteria of model preference normally used, such as cross validation tests or others as AIC (Akaike 1973), BIC (Schwarz 1978) and DIC (Spiegelhalter *et al.* 2002)? How do the answers to these questions affect the way genomic selection models are constructed and compared, especially regarding the choice of model covariates?

The rationale applied here to tackle these questions could be summarized by the following steps: 1) reviewing that association and causation are different types of information, and that inferring causal effects and associations are different tasks given that causal assumptions are required by the former, 2) deducing with causal graphs that inferred genomic predictors are only useful for selection if they reflect causal genetic effects, 3) concluding from 1) and 2) that the construction of genomic selection models for inference of breeding values should obey the requirements of causal inference, while failing in doing so may result in obtaining genomic predictors that are poor representations of the genetic effect (regardless of the resulting predictive ability), 4) providing examples to this conclusion by showing with simulation studies that usual model comparison criteria may not reflect the quality of breeding values inferences from fitting alternative models with different sets of covariates, and 5) suggesting that criteria for the identifiability of causal effects with respect to specific causal assumptions are theoretically more suitable for model construction/comparison in the context of genomic selection.

For the presentation of the aforementioned rationale, the manuscript is structured as follows: we present models normally used to infer genomic breeding values in the section **Standard models applied to genomic selection**. In **Graph-theoretic terminology**, we present the graph concepts necessary to attain the article goals. In **Effect x association**, we stress pertinent differences between learning causal and statistical information from observational data. In **The genetic effect**, we demonstrate that selection relies on a causal effect of genotypes on a trait, and that this is the relevant information for selection programs rather than any other source of association between genotypes and phenotypes that could also enable genomic predictions. In **Simulated data examples**, we illustrate the points made in the last section by showing that the choice of model covariates has an important and non-trivial influence on the inference of the genetic effects, and that criteria usually applied for model comparison may not indicate if a specific choice is suitable. In **Identifiability criterion for causal effects inference**, we propose using covariate choice criteria for causal inference that have not been used for inferring genetic effects. Final remarks are presented in **Discussion**.

**Standard models applied to genomic selection**

The inference of genetic effects involves fitting mixed models as

$$y_i = \mathbf{x}_i \boldsymbol{\beta} + u_i + e_i. \tag{1}$$

In this model, $y_i$ is a phenotype for a trait recorded in the $i^{th}$ individual. The model express the phenotypic value as a function of other variables in the right-hand side, including model residuals $e_i$, breeding values or additive genetic effects $u_i$ and other covariates in $\mathbf{x}_i$.

The row vector $\mathbf{x}_i$ traditionally accommodates information of variables other than genetic effects included in model's right-hand side. They may include environmental variables or phenotypic traits different from the one to which $y_i$ is related. The column vector $\boldsymbol{\beta}$ contains fixed "effects" in a frequentist sense. The last term is presented between quotes to stress the difference between a term frequently used to label the entries in $\boldsymbol{\beta}$ and the formal meaning attributed to the term *effect* throughout this manuscript.

Inference of individual breeding values depends on assigning to their predictors a covariance structure represented by a genetic distance matrix. Traditionally, this covariance structure is derived from pedigree distances (Henderson 1975). As information on genome-wide marker genotypes became more available, one alternative is to substitute the pedigree-based distance matrix to a genomic distance matrix that is fully (VanRaden 2008; de los Campos 2009a) or partially (Legarra *et al.* 2009) computed from these marker genotypes. In the genomic context, the genetic effect term could be substituted to an explicit function of marker genotypes and marker specific coefficients, so that the model would become:

$$y_i = \mathbf{x}_i\boldsymbol{\beta} + \mathbf{z}_i\mathbf{m} + e_i \qquad [2]$$

where $\mathbf{z}_i$ is a row vector containing information on genotypes for different SNP (single-nucleotide polymorphism) markers recorded on the $i^{\text{th}}$ individual, and $\mathbf{m}$ is a vector of marker additive "effects". Many different modeling approaches have been proposed under this framework, such as Bayes A and Bayes B (Meuwissen *et al.* 2001), Bayes C (Habier *et al.* 2011), Bayesian Lasso (Park and Casella 2008; de los Campos *et al.* 2009b) and many others. The essential difference between many of these models is the prior distribution assigned to the marker coefficients (de los Campos *et al.* 2013a; Gianola 2013).

**Graph-theoretic terminology**

The structure of how variables are causally related can be represented by a directed graph (Pearl 1995; Pearl 2000) such as the one depicted in Figure 1. A directed graph consists on a set of nodes (representing variables) connected with directed edges (representing pairwise causal relationships). A pair of connected nodes as $a \rightarrow c$ means that $a$ has a direct causal effect on $c$. However, causal effects can be indirect such as the effect of $b$ on $e$ through $c$ ($b \rightarrow c \rightarrow e$). Any sequence of connected nodes where each node does not appear more than once is called a path (e.g. $d \leftarrow a \rightarrow c \leftarrow b$, $a \rightarrow d \rightarrow e$). In a path, a collider (Spirtes *et al.* 2000) is a node towards which arrows are pointed from both sides (e.g. $c$ in $d \leftarrow a \rightarrow c \leftarrow b$). Paths can potentially transmit dependence between the nodes on the extremes (active paths). Otherwise, they can be blocked (inactive paths), and therefore transmit no dependence. Marginally, non-colliders allow the flow of dependence. For example, in *a→d→e* there is dependence between *a* and *e*. This is suitable given the causal meaning of this path, since *a* should be dependent of *e* since *a* affects *e*, and *d* allows the flow of dependency since it mediates the causal relationship. Likewise, *d* and *c* are expected to be dependent through the path *d←a→c*. This is expected given that variable *a* is a common influence on both of them, and therefore, *a* allows flow of dependence as well. On the other hand, a collider is sufficient to block a path. For example, in *a→c←b*, *c* is just a common consequence of *a* and *b*, and this do not imply in dependence between this pair (i.e. observing a value for *a* do not change the expected value for *b* just on the basis of having a common consequence in *c*). Upon conditioning, these properties of colliders and non-colliders are reversed. This means that conditioning on non-colliders block the path. For example, in *a→d→e*, conditionally on knowing the value of *d*, learning the value of *a* does not give any additional information about *e*. The same goes for *d←a→c*. Contrariwise; conditioning on colliders turns it into a node that allows the flow of dependence. For example, in *a→c←b*, once the value of *c* is known, then observing the value for *a* updates the expected value for *b*. Paths that present marginal flows of dependence either represent a causal path (e.g. *a→d→e*) or back-door paths (e.g. *d←a→c→e*). The latter are marginally active paths containing both extreme nodes with arrows pointed towards them, and represent a relationship between these pair of nodes that is not causal, but it is a source of association.

    Notice that a graph is a good way to encode causal information and assumptions. However, it only provides a qualitative representation of causal relationships, and therefore it does not sufficiently specify a causal model. For example, from *a→c* it is not possible to deduce the magnitude of the effect, or its sign (positive or negative), and therefore this is not sufficient

to determine the resulting joint probability distribution. A causal graph can be interpreted as a family of causal models from which those qualitative causal relationships can be deduced. However, by exploring the *d*-separation criterion (Pearl 1988; Pearl 2000), graphs are very efficient in representing the conditional independences among variables that necessarily follow from the causal information they encode. Two nodes are *d*-separated in a graph conditionally on a subset of the remaining nodes if there are no active paths between them under this circumstance. For example, *a* and *e* are *d*-separated conditionally on *d* and *c*, as both paths between these two variables (*a*→*d*→*e* and *a*→*c*→*e*) become inactive in this context. This means that in the joint probability resulted from any causal model with the given structure, *a* and *e* are independent conditionally on *d* and *c*. Notice however, that conditioning on only one of either *c* or *d* is not sufficient for *d*-separation, as one of the paths between *a* and *e* becomes active. Likewise, *b* and *d* are marginally d-separated as both paths between them contain a collider (i.e. *d*←*a*→*c*←*b* and *d*→*e*←*c*←*b*), but they are not *d*-separated conditionally on *e* or on *c*.

**Effect x association**

There is a sharp distinction between statistical/associational and causal concepts (Pearl 2000; Spirtes *et al.* 2000; Pearl 2003; Rosa and Valente 2013). The first group of concepts can be completely described from a joint probability distribution alone. Examples of such concepts are correlation and regression coefficient. On the other hand, causal concepts involve deeper features of the relationship between variables: it concerns structural aspects of the data generation process, allowing one to predict how the joint probability of events would change in case of external interventions. Causal relationships among variables result in a joint distribution as an observational consequence, and this distribution may reflect some features of these underlying causal relationships, but not sufficiently for defining them completely. Thus, causal information cannot be deduced from a joint distribution alone. One example of causal concept is the effect between two variables.

Consider two variables $x$ and $y$. If it is said that $x$ have an effect on $y$, it is implied that the value of $x$ have bearings on the process that generates the value of y. This conveys an asymmetric relationship that is made more transparent by the concept of external intervention: From "$x$ have an effect on $y$", it can be deduced that intervening on the value of $x$ changes the expected value of $y$, while nothing can be implied about changes in the expected $x$ from interventions on $y$. One observational consequence of a causal relationship between variables is an association (which is a statistical concept). However, different causal relationships could result in the same pattern of association. As a simple example, the following four hypotheses are equally compatible with an observed association between $x$ and $y$: a) $x$ affects $y$ (i.e. $x \rightarrow y$), b) $y$ affects $x$ (i.e. $x \leftarrow y$), c) both $x$ and $y$ are affected by a set of variables Z (i.e. $x \leftarrow Z \rightarrow y$) and d) any combination of the previous three hypotheses. As different hypotheses are equally supported by a given association (or distribution), this information is not sufficient to learn the magnitude of a specific causal effect without further assumptions.

Still regarding the same scenario, consider fitting a simple linear regression model $y_i = \mu + x_i \beta + e_i$. In the context of a bivariate normal distribution, the linear regression coefficient $\beta$ could be defined in terms of covariances and variances, all of them defined from the distribution. By itself, this regression coefficient does not have any causal content, as it just describes the change in expected $y$ if the value *observed* for $x$ was larger by one unit. In this context, it carries no information about how the value of $y$ would change from increasing the value of $x$ through intervention. If instead, the model was written as $x_i = y_i \alpha + e_i$, than $\alpha$ would reflect the same association expressed by $\beta$, but possibly with a different scale. For $\beta$ to be called an effect, (i.e. $\hat{\beta}$ identifies the change in expected *y* if *x* was physically increased by one unit), causal assumptions are necessary. In this example, it is required to assume that both variables do not have a common cause and that *y* does not have any effect on *x*.

Assume a causal relationship between *x* and *y* that includes a third variable that affects both of them as presented in Figure 2a. In this case, $\hat{\beta}$ obtained from fitting $y_i = x_i\beta + e_i$ does not represent an effect. The coefficient $\hat{\beta}$ represents the marginal association between *x* and *y* that is given by an unspecified combination of two active paths contributing to its magnitude: *x*→*y* and *x*←*z*→*y*. The combinations of causal effects contained in both paths that would result in the same $\hat{\beta}$ involve infinite possible values for the effect *x*→*y*. This illustrates how learning the causal effect *x*→*y* from $\hat{\beta}$ is confounded by the back-door path *x*←*z*→*y*. Notice however that if the goal is purely predicting *y* from *x*, then knowing $\hat{\beta}$ is sufficient. On the other hand, as conditioning on *z* blocks (or inactivates) the confounding path, a suitable alternative would be to infer the effect from a model where $\hat{\beta}$ captures the association between *x* and *y* conditionally on *z*, such as from $y_i = x_i\beta + z_i\alpha + e_i$. By applying this model, and under the causal assumptions depicted in Figure 2a, $\hat{\beta}$ could be declared to represent the effect of *x* on *y*.

Including covariates is not always beneficial for causal inference. If a variable *w* is assumed to be affected by both *x* and *y* as depicted in Figure 2b, then conditioning on it activates the path *x*→*w*←*y*. For this reason, this non-causal source of association would contribute to $\hat{\beta}$ if it is estimated from $y_i = x_i\beta + w_i\alpha + e_i$. Accordingly, this model could be still valid for predicting *y* from *x* and *w*, but including *w* as a covariate confounds the estimation of the effect between *x* and *y* from $\hat{\beta}$. Nevertheless, the assumed causal relationship between *x* and *y* allows identifying the causal effect from $\hat{\beta}$ obtained by fitting $y_i = x_i\beta + e_i$.

Formal criteria for choosing covariates aiming for the identifiability of causal effects are given by the back-door criterion (Pearl, 2000), or more generally by the criterion described by Shpitser *et al.* (2012). The back-door criterion is presented in more details in the section **Identifiability criterion for causal effects inference**. The main message here is that only after assuming a specific qualitative causal relationship among variables (which could be represented by a graph) and after verifying whether a coefficient $\hat{\beta}$ of a given model identifies the effect of interest accordingly to the causal assumptions, then $\hat{\beta}$ can be declared to express the effect of *x* on *y*. Conversely, simple inferences of regression coefficients are associational information and do not require any of such assumptions. However, they cannot be interpreted as having a valid causal meaning. This is a remarkable difference between a regression study and a causal inference.

**The genetic effect**

In basic quantitative genetics settings from which [1] and [2] were derived, phenotypes are represented simply as:

$$y = g + e,\qquad\qquad\qquad [3]$$

where *g* can be viewed as a function of the genotype of an individual and *e* accounts for environmental factors (Falconer 1989). The expected phenotypic value associated with a given genotype is called genotypic value. In the context of animal and plant breeding, models for genetic evaluation generally assume the signal between genotypes and phenotypes as additive, in which case it could be called "breeding value" or "additive genetic effect" (here represented as $u$). In this context, predictors obtained on the basis of genomic information aim towards capturing this additive signal. These predictors customarily have a defining role in selection decisions. The same applies to pedigree-based predictors, but in this manuscript we focus on the genomic selection context. The signal between genotype and phenotype will be assumed as additive from now on.

One question posed in **Introduction** is if the inference of these genomic predictors should be conducted as inferences of causal effects. If the usefulness of these predictors for selection does not rely on a causal meaning, then the job should be conducted under the familiar framework of the regression analysis. Contrariwise, if usefulness of such predictors depends on the ability to fit a causal signal, then the inference of breeding values undeniably belongs to the field of learning causal information from observational data, and the extra requirements for this type of analysis should not be ignored. If the latter is accepted, one should avoid interpreting predictors obtained from fitting genomic selection models as genetic effects without regard to the causal assumptions that grant identifiability to this causal information. The decision about treating genomic predictions for selection as causal inference or as standard regression studies is not straightforward since the available literature provides information that seems to support both alternatives in different occasions, as presented next.

Some of the labels usually attributed to the quantity in question indicate that it should be treated as an effect (e.g. genetic *effect*), while other labels leave room for doubts (e.g. breeding value, expected progeny difference). Authors in quantitative genetics present the concept of breeding values mostly under a causal rather than under an associational framework, as their definitions for it employ terms as "effect", "causes of variability", "influence", "contribution", "transmission of values", and so forth; see for example (Fisher 1918; Falconer 1989; Lynch and Walsh 1998). On the other hand, the literature on animal breeding indicates that the regression approach is deemed as suitable. For example, recent literature on the development of methods for genomic prediction might inspire some skepticism about accounting for or learning biological architecture information from observational data. As the reasoning follows, such architecture involves extremely complex pathways and interactions, and suitable learning would require much more data than what is available. One possible implication is that genomic selection models should be focused on obtaining relevant predictions rather than being used to learn architectural information. The avoidance of accounting for or learning architectural features might be interpreted as a departure from the context of causal inference. This interpretation may take place since the indispensable causal assumptions are basically structural/architectural information that needs to be brought in so that inferences could be interpreted as causal. Actually, the very output of causal inference is architectural information that goes deeper into the relationship between two variables (such as genotypes and phenotypes) than what is typically explored for pure predictions. Furthermore, even the nomenclature recently applied to predictors in genomic selection studies indicates a departure from the causal context (e.g. genome-enabled prediction, whole genome-regression). Finally, while the focus on predictive ability and other statistical aspects of model construction and comparison is strongly established in this area of research, the discussion on the pitfalls and challenges of causal inference is virtually nonexistent in the literature related to genetic prediction for selection purposes. This historical focus could be seen as an indication that causality is not the matter in animal and plant breeding. The exceptions are models used explicitly to convey causal relationships between variables, as Structural Equation Models (Gianola and Sorensen 2004; Rosa *et al.* 2011; Valente *et al.* 2013). However, even in the applications of these models the discussion is never focused on the identifiability of causal effects from breeding value predictors.

The decision on treating the inference of genomic predictors as a regression analysis or a causal inference is not the same as deciding if there is an effect between genotype and phenotype. In other words, one should not decide for the causal inference approach only because the genotype is believed to affect phenotype. This possibility would not be hard to accept in most cases, but this is not important since the regression approach does not assume the absence of such relationship. The defining point is verifying if breeding programs goals depend on learning causal information or if the pure ability to obtain predictors from genotypes is sufficient for them. In general, learning causal information is required when one needs to infer how a set of variables is expected to respond to external interventions. Specifically for the issue here studied, the decision on the analysis approach depends whether learning the effect of an intervention is necessary for selection. A basic structure involved in selection could be represented as in Figure 3a, where $G$ is a node that represents a whole-genome genotype for some individual, and $y$ represents a phenotype for some trait recorded from it. For this point, consider that there is a signal between $G$ and $y$ but the causal relationship that is its source is not resolved, so that it is represented by an undirected edge. Also, let $G'$ be the genotype of an individual pertaining to a future generation, such as an offspring of the individual with genotype $G$. $G'$ is connected to its own phenotype $y'$ in a similar fashion of $G$ and $y$. Here, $G$ and $G'$ are connected with an edge directed towards $G'$, as genotype $G'$ is affected by $G$ through inheritance.

Selection programs involves modifying the genotypes of individuals of the next generation ($G'$), and this modification is expected to result in shifting the phenotype $y'$ as a response. Therefore, selection relies on a causal relationship directed from $G$ to $y$, such as given in Figure 3b. Typically, the modifications attempted consist on increasing the frequency of alleles with desirable effects on a phenotypic trait. Such alleles are inherited from parents (*G*). Therefore, selecting the best parents depends on identifying individuals with genotype *G* carrying alleles with the best *effects* on phenotype *y* (i.e. it involves identifying genotypes with best *effects* on phenotypes). This is different from identifying individuals with alleles associated with the best phenotypes (i.e. identifying genotypes associated with best phenotypes). The genomic predictors should identify a causal effect of *G* on *y* to be relevant for selection. Notice that the assumption that *G* affects *y* (which is not hard to accept for most cases) is not sufficient to guarantee this identifiability. This might not be clear when the causal relationships assumed are as in Figure 3b because in this case the magnitude of the effect of *G* on *y* is identified by the marginal association between them (i.e identifying genotypes marginally associated with best phenotypes is the same as identifying genotypes with best effects on phenotypes). However, differentiating effect from the association is important because other sources of association may occur, as discussed ahead. Additionally, spurious associations could even be created by bad modeling decisions. At any rate, this indicates that here as for any causal inference, interpreting predictors as genetic causal effects involves assuming a specific causal relationship between *G* and *y* and proposing a model that allows identifying this effect from the other possible sources of associations. This approach is required even in the simpler context with no sources of confounding. Genetic predictors obtained from fitting a model as simple as [3] cannot be declared as genetic effects if a causal relationship between *G* and *y* such as in Figure 3b is not assumed.

To illustrate these concepts, consider a scenario where *y* is not affected by *G* (i.e. *y* is not effectively heritable), but phenotypes of relatives tend to be similar to each other due to a common environment effect (i.e. *G* and *y* are associated). According to the common-cause assumption (Reichenbach 1956), two variables can be declared as affected by a third variable if they are dependent but they do not affect each other. As *G* and *y* do not affect each other even though they are associated, this assumption is summoned here to represent the relationship between *G* and *y* with a double headed arrow (Figure 3c). Such arrows summarize a back-door path representing the common cause. As in this case *G* and *y* are associated, predictions of phenotypes from genotypes (i.e. genome-enabled prediction, or whole-genome regression) could still be attained. However, trying to improve *y'* from modifying *G'* would be useless as *G'* have no causal effect on *y'*. Notice that a genomic predictor obtained under this scenario would capture this non-causal signal, and for this reason could not be properly called as a predicted genetic effect.

Consider another scenario (Figure 3d) where the observed signal between *G* and *y* is due to a combination of causal and spurious sources. The response of *y'* to interventions on *G'* would only depend on the causal path between both variables. Distinguishing the association generated from this causal path from the spurious ones would be important for selection. This is necessary to distinguish between the genotypes with best effect on *y* from the genotypes associated with the best *y*'s, and therefore to discriminate the best breeders. Again, all these issues are not relevant when one intends just to predict a future performance instead of selecting individuals for breeding. In this case any signal could be explored regardless of its sources (e.g. a combination of causal effects and spurious associations). This scenario additionally illustrates that assuming that *G* affects *y* is not sufficient for declaring that genome-based predictors represent breeding values.

The above formulations indicate that predicting genetic effects cannot be treated as a standard regression problem. It should be noticed that identifiability of causal information is distinct from likelihood identifiability. For example, from a sample of bivariate normal distribution for two variables *x* and *y*, the linear regression coefficient in the model $y_i = \mu + x_i\beta + e_i$ is identifiable from the likelihood function if the sample of observations presents two or more distinct values. However, the identification of an effect from inferred $\hat{\beta}$ is not attainable (even with infinite data points) if the assumed causal relationship between *x* and *y* is as Figure 2a.

A simple numeric example of these concepts would consist of two genotypes $G_A$ and $G_B$, marginally associated with expected phenotypic values 2 and 3, respectively. This information is useful by itself for a "genomic" prediction. For example, one could declare that if a genotype *observed* for some individual was equal $G_B$, then the expected phenotypic value would be one unit larger than if the *observed* genotype was $G_A$. This would be equally valid under any of the structures presented in Figure 3, so no causal assumptions are required. On the other hand, interpreting the aforementioned association as an increase in the expected phenotype by one unit if an individual with genotype $G_A$ had it *changed* to $G_B$, then it would be necessary to assume that this association reflects a causal effect with no confounding. This is equivalent to assuming the structure in Figure 3b. It should be stressed that in essence, both situations requires predicting phenotypes, but the latter involves predicting phenotypes under an ideal intervention on the genotype. Changing an individual's genotype is not a feasible technique, but deciding between two candidates for breeding with genotypes $G_A$ and $G_B$ would result in physically changing genotypes in future generations, and the prediction on how phenotypes will respond from using $G_A$ or $G_B$ as breeders depend on predicting how each genotype affects the phenotypes.

The basic setting of quantitative genetics as presented in [3] contains only the genotypes and the analyzed trait as variables explicitly considered in the model. As explained, identifying genetic causal effects from fitting genomic predictors is not an issue in this simplified context. However, the models applied to field data generally incorporate further covariates. In dealing with causal inference from fitting a model, the decision of including covariates should be done as to achieve the identifiability of the relevant information according to causal assumptions. Remarkably, the decisions on model construction for breeding values inference are predominantly molded by statistical criteria, such as significance of associations, goodness of fit scores or model predictive performance, while requirements for identifiability of causal effects are largely ignored. This is an important issue, because both including and ignoring covariates may produce good predictors of phenotypes that are bad predictors of genetic effects. In the next section we provide examples of how statistical criteria may not be a good guide for models construction when the goal is the inference of breeding values.

**Simulated data examples**

In this section, we provide examples based on simulated genotypes and phenotypes. The R (R Development Core Team 2009) script used for such simulation was adapted from Long *et al.* (2011). Genome consisted of 4 chromosomes with 1 Morgan each, 15 QTL per chromosome and 5 markers between consecutive pairs of QTL (320 marker loci). An initial population of 100 individuals (50 males and 50 females) was considered with no segregation. Polymorphisms were created after 1000 generations of random mating and a probability of 0.0025 of mutation for both markers and QTL. The number of individuals per generation was maintained at 100 until generation 1001, when the population was expanded to 500 individuals per generation. Random mating was simulated for 10 more generations. Data and genotypes for the individuals of the last four generations (2000 individuals) were used for the analyses. Four different scenarios were studied. For each one, traits were simulated with distinct sampling parameters and different relationships between traits, which will be specified later on.

Data was analyzed via Bayesian inference on the basis of model [2]. For each scenario, there was a variable that could either be included as a fixed covariate in $\mathbf{x}_i\boldsymbol{\beta}$ or ignored. Therefore, two alternative models were fitted for each scenario, differing only on the construction of $\mathbf{x}_i\boldsymbol{\beta}$. These two models are referred to as model c and model ic (c standing for "covariate", and ic standing for "ignoring covariate"). Traditionally, in mixed models it is common to consider not only environmental variables, but also phenotypic traits as fixed covariates. For example, a model for studying age at first calving or a behavior trait may "correct for" or "account for" body weight at a specific age; a model for somatic cells score may account for milk production; a model studying first calving interval may account for age at first calving and so forth. Here we evaluate simple scenarios with only two alternative models, but applications on field data may involve much larger spaces of models.

To fit these models, the R method BLR (de Los Campos *et al.* 2013b) was used. Based on [2], it follows that $p(y_i | \boldsymbol{\beta}, \mathbf{m}) \sim N(\mathbf{x}_i\boldsymbol{\beta} + \mathbf{z}_i\mathbf{m}, \sigma_e^2)$, and the prior distribution assigned to parameters is:

$$p(\boldsymbol{\beta}, \mathbf{m}, \sigma_m^2, \sigma_e^2) = p(\boldsymbol{\beta}) p(\mathbf{m} | \sigma_m^2) p(\sigma_m^2) p(\sigma_e^2)$$
$$\propto N(\mathbf{0}, \mathbf{I}\sigma_m^2) \chi^{-2}(df_m, S_m) \chi^{-2}(df_e, S_e)'$$

where an improper uniform distribution was assigned to $\boldsymbol{\beta}$; $N(\mathbf{0}, \mathbf{I}\sigma_m^2)$ is a multivariate normal distribution centered in $\mathbf{0}$ and with diagonal covariance matrix $\mathbf{I}\sigma_m^2$, where $\mathbf{0}$ is a vector with zeroes and $\mathbf{I}$ is an identity matrix, both with appropriate dimensions; $\chi^{-2}(df_m, S_m)$ and $\chi^{-2}(df_e, S_e)$ are scaled inverse chi-square distributions specified by degrees of freedom $df_e = df_m = 3$ and scales $S_m = 0.01$ and $S_e = 50$. Therefore, parameters of the last two distributions were treated as known hyperparameters.

The predictive ability was evaluated as usually to compare models for genomic prediction. Here, we have performed 10-fold cross validations and evaluated the predictive ability according to the correlation between $y_i$ in the testing set and $\hat{y}_i = \mathbf{x}_i \hat{\boldsymbol{\beta}} + \mathbf{z}_i \hat{\mathbf{m}}$, which is a prediction conditional on observed values for $\mathbf{x}_i$ and $\mathbf{z}_i$ in the testing set, and posterior means $\hat{\boldsymbol{\beta}}$ and $\hat{\mathbf{m}}$ obtained from the training set. The predictive performance was also evaluated using the correlation between the genomic prediction $\mathbf{z}_i \hat{\mathbf{m}}$ and the phenotype in the testing set corrected for fixed "effects" $y_i^* = y_i - \mathbf{x}_i \hat{\boldsymbol{\beta}}$. This evaluates the ability to predict deviations from fixed effects. As genetic effects themselves can be viewed as deviations from fixed effects, this last test can be judged as more suitable when the goal is predicting breeding values. Finally, we have also evaluated models according to the correlation between genomic predictors $\mathbf{z}_i \hat{\mathbf{m}}$ and the true genetic effect $u_i$, which is the relevant information for selection purposes. Here we intend to demonstrate that cross validations, even if aiming to evaluate the ability to predict deviations from fixed effects, may not indicate the model that best provides the relevant information for selection. In the next section, based upon recognizing that the problem requires causal inference, we propose deciding on covariate inclusion/exclusion based on criteria for causal effects identifiability.

In the first simulation scenario, $y_1$ is generated as a trait that is not genetically influenced. However, it affects a heritable trait $y_2$, such as depicted in Figure 4a. The recursive model used to sample data is presented as a mixed effects Structural Equation Model (Gianola and Sorensen 2004; Wu *et al.* 2010; Rosa *et al.* 2011) in Figure 4b. The trait $y_1$ could be thought of, for example, as the liability of a non-heritable disease that affects a heritable trait as milk yield. Here, consider one is interested in genetic evaluations for $y_1$. One possible model for that task includes $y_2$ as a covariate (model c), as for example a genomic selection model for incidence of a disease that "corrects for" milk yield. An alternative model ignores $y_2$ (model ic). In Figure 4c, we can observe that the criteria normally used to compare models suggest that model c is the best model, for it predicts phenotypes more accurately and provides better predictions of random deviations from expected phenotype given fixed effects. Furthermore, it granted more variability to the genomic predictors ($\mathbf{z}_i\hat{\mathbf{m}}$), so that they explain a reasonable proportion of the variability of $y_1$. On the other hand, model ic provides poor predictive ability from genomic information. However, if one is interested in selection, the latter is actually the best model, because the genetic predictors better reflect the genetic causal effects, or in this case, their absence. Model c provides better performance on cross-validation tests, but interpreting its predictors as reflecting genetic effects suggests that $y_1$, which is actually non-heritable, would respond to selection. This confusing result comes about because, in this model, the genomic predictor apprehends the signal between the genome-wide genotype and $y_1$ conditionally on $y_2$. Conditioning on a common consequence of both G and $y_1$ constitutes conditioning on a collider of the path G$\rightarrow y_2 \leftarrow y_1$, which activates it. This creates a non-null signal between the markers and $y_1$ that do not reflect a causal effect. On the other hand, the model that ignores $y_2$ does not create a spurious association, as the genomic predictors explore the marginal associations between the genotype and $y_1$, which is null, reflecting the absence of effect.

Consider an alternative scenario which is similar to the previous one, but assigning variable genetic effects to $y_1$ (Figure 5a and 5b). Consider we are still interested in selection for $y_1$ and have the same alternative models for inferring breeding values: one including $y_2$ as a fixed covariate (model c) and other ignoring it (model ic). Observe that in this scenario, $y_1$ could potentially respond to selection, but optimization of this response would depend on the accuracy in inferring genetic causal effects. As in the last scenario, model c provides the best predictive ability according to $cor(y_1, \hat{y}_1)$ and $cor(y_1^*, \mathbf{z}\hat{\mathbf{m}})$, as depicted in Figure 5c. However, the correlation between predicted genetic effect and true genetic effects on $y_1$ ($cor(u_1, \mathbf{z}\hat{\mathbf{m}})$) indicates that model ic better identifies the target quantity. For this scenario, there is no marginally active path between $G$ and $y_1$ aside from $G \rightarrow y_1$. This causal connection is the only path that contributes to the marginal association between $G$ and $y_1$, which is explored by model ic. The path $G \rightarrow y_1$ also contributes to genetic predictors of model c, but a second source of association is introduced due to including $y_2$ as a covariate, activating $G \rightarrow y_2 \leftarrow y_1$. As a result, the signal apprehended by predictors corresponds to a combination of both active paths. The contribution of this non-causal signal improves predictions evaluated by cross-validations, but harms the ability to identify the genetic causal effect. On the other hand, predictors from model ic are not confounded for representing genetic effects, even though it does poorer on cross-validation tests.

In a third scenario (Figure 6a and b), let the causal structure be similar to last scenario, but now the interest is studying $y_2$ instead of $y_1$. Two alternative models involves including $y_1$ as covariate (model c) or ignoring it (model ic). Notice that the target quantity $u$ is not what is represented by $u_2$ in Figure 6b. The variable $u_2$ represents only the genetic effects on $y_2$ that are not mediated by $y_1$, but genetics also affect $y_2$ through $G \rightarrow y_1 \rightarrow y_2$. The response to selection on a trait in general depends on the overall effect of the genotype on that trait, regardless if effects are direct or not. Therefore, the target of inference here is not $u_2$, but $u = 0.8u_1 + u_2$. Here again, standard cross validation results indicated that the model that achieves worst prediction of target genetic effects (model c) should be preferred. As mentioned, the target quantity consists on a combination of causal effects through two paths, but including $y_1$ as a covariate blocks one of the paths in the signal captured by $\mathbf{z}_i \hat{\mathbf{m}}$ in model c. As a result, just part of the causal effect sought is fitted. On the other hand, model ic does not block this path and the overall effects are better identified by genomic predictors.

One extra issue that can be stressed in this example involves the use of the variability of genetic predictors to compare models. As the justification goes, if a model infers larger "genetic variance" than other models, this indicates an ability to capture a larger proportion of the true genetic variability. It is implied that the larger the inferred "genetic variance", the better inferred predictors represent true genetic effects. This example illustrates that this may not be necessarily true (see the same applying to examples in Figures 4 and 7). Furthermore, it would be expected that a model that blocks part of the genetic causal effects would result in less genetic variability expressed, at least in a simpler scenario where direct genetic effects are independent. However, this is not the case if direct genetic effects on traits are negatively associated and the causal effect between traits is positive (as applied in this simulation scenario), or vice-versa. In this case, blocking one causal path may increase the variability of the predictors. However, they should not be blocked when the target is the overall effect, even supposing it is given by the combination of two "antagonist" causal paths.

Consider a fourth scenario (Figure 7a and b) where the response phenotypic trait $y$ is affected by $G$ but there is an extra source of marginal association between genotype and phenotype, and there are measurements for a third variable $X$ that lies in this path. For the simulation, we emulated a setting where a trait $y_1$ is weakly affected by the genotype, and it is also affected by an environmental factor $X$. We considered however that $X$ is associated with the individual genotypes through a back-door path. One example for such relationship would be a study involving beef cattle, where categories in $X$ are assigned to different farms, and the best farms (i.e. the farms with the best effects on $y_1$) tend to buy semen from bulls with the best genetic effects for a different trait (e.g. the best genetic effects on $y_2$). In this case, genotypes of parents of each $i^{th}$ individual affect the $X$ category that will influence their offspring phenotype. Additionally, genotypes of parents affect the offspring's genotype, constituting a back-door path between $G$ and $X$. In Figure 7a, such path is represented by a double-headed arrow connecting this pair of nodes. Accordingly, four farm effects were simulated, so that four categories were assigned to $X$. Also, genetic effects on $y_1$ and $y_2$ were negatively correlated. Because of the negative genetic correlation, the farms with larger effects for $y_1$ tended to present individuals with worse genetic effects on the same trait. It should be stressed that although this scenario results in a causal effect of the genotype of a bull on its offspring phenotype through $X$, this is not an effect that a breeding program (even on a sire model perspective) should explore. Two alternative models would be including farm as a categorical covariate (model c) or ignoring it (model ic). From the cross-validation results (Figure 7c), notice that although model c provides better predictions of $y_1$ in the testing set, model ic is much better at predictions of $y_1^*$ in the same set. Additionally, fitting model ic apparently results in an reasonable apprehension of genetic variability of $y_1$, unlike model c. However, the correlation between predictions and target value $u_1$ demonstrates that including the covariate *X* results in better identification of the genetic effects. The variability of predictors obtained from this model correctly suggests that response to selection would be slow. Model ic seems to suggest good ability to predict genetic effects and also a reasonable genetic variability. However, because of the negative correlation between the genomic predictor and the target effect, using such predictors for selection decisions is expected to result in negative selection for the trait of interest.

These examples illustrate that, in essence, traditional methods used for model comparison do not evaluate the quality of the prediction of the genetic causal effects. It is not implied that they always point towards the worst model. Of course, in many other instances with different structures and parameterizations, these comparison methods would eventually point toward a suitable model. However, the simulations were used as *exempla contraria* to show that pure genomic predictive ability is not the point. The ability to predict is not sufficient to judge a model as useful for selection. The effective criterion to recognize a model as useful for this context cannot even be properly articulated under the prediction or even statistical framework.

**Identifiability criterion for causal effects inference**

In most settings, it is the overall effect of *G* on *y* that should constitute the inference target for selection purposes, even though this overall effect could eventually be disentangled into direct and mediated effects. To verify if a model is able to identify the target causal effect given the assumed causal relationships among variables, one could apply the so-called back-door criterion (Pearl 2000):

For any two variables $x$ and $y$ in a causal diagram **D**, the total effect of $x$ on $y$ is identifiable if there is a set of measurements **Z** such that:

1 – No member of **Z** is a descendant of $x$; and

2 – **Z** $d$-separates $x$ from $y$ in a subgraph $\mathbf{D}_{\underline{X}}$ formed by deleting from **D** all arrows emanating from $x$

Moreover, if the two conditions are satisfied, then the total effect of $x$ on $y$ is given by the fitted coefficient $\beta_{yx.\mathbf{Z}}$, which is a regression coefficient of $y$ on $x$ conditionally on **Z**.

In Figure 8 we present some simple examples of applications of the back-door criterion for constructing linear models that identify the effect of $x$ on $y$ depending on causal relationships assumed. Violating the first rule of the criterion (i.e. including a descendant of $x$ as a model covariate) may either block causal effects by inactivating a causal path or create associations that are non-causal by conditioning on a collider. Additionally, violating the second rule may keep back-door paths active, and non-causal paths may end up contributing to the effect estimator.

The application of such criterion in predicting breeding values means that one should articulate a priori how the involved variables (genotypes, response phenotypes and covariate candidates) are assumed to be structurally articulated. The inclusion of a covariate in a genomic selection model is analogous to including this variable in the set **Z** presented in the definition of the back-door criterion. The signal between $G$ and $y$ captured by genomic predictors is expressed conditionally on the covariates included. The requirement of prior architectural information for covariate choice is one apparent disadvantage of this approach. However, as demonstrated ahead, the information required may not constitute too detailed descriptions of biological pathways, but simple and general biological knowledge. For this reason, necessary assumptions may not be hard to accept. Nevertheless, one should keep in mind that regardless if sufficient causal assumptions are difficult to accept in specific scenarios, selection decision still depends on causal information. Such information is not provided from the standard regression/prediction approach by itself, regardless of how easier it is to accept its assumptions. The back-door criterion would be sufficient to compare models in the examples given in the last section as described next.

In both scenarios depicted in Figures 4 and 5, the goal was to infer causal effects from genotypes to $y_1$. Model c and model ic would correspond to $\mathbf{Z} = \{y_2\}$ or $\mathbf{Z} = \{\ \}$, respectively. The back-door criterion would indicate that ignoring $y_2$ is required if this variable is assumed as affected by the genotype. In this case, model c violates the criterion. As there are no additional paths or known backdoor confounders, model ic meets the criterion and would be preferred even though predictive ability tests suggests otherwise. Notice also that the prior causal information required is simply assuming $y_2$ as heritable. Consider the hypothetical setting where the goal is to predict genetic effects for the incidence of a disease while having the option to include milk yield as a model covariate. Regardless of the results from cross-validation studies, recognizing the inference of genetic effects as a causal inference and applying the back-door criterion would point towards the model that ignores milk yield information. The causal assumption that supports that decision is simply acknowledging that milk yield is heritable, which is not hard to accept

The basis for the comparisons involved in these two scenarios would be similar to what applies to the scenario presented in Figure 6. The back-door criterion indicates that model ic should be preferred. However, the underlying consequence of conditioning on some other heritable trait would be different. For the settings of Figures 4 and 5, the consequence would be creating an association that does not correspond to an effect. In the scenario depicted in Figure 6, the consequence of including $y_2$ in the model would be blocking part of the effect to be inferred. Notice that discerning the type of mistake depend on knowing how traits are causally related, which may be difficult to learn in some settings. Fortunately, the correct modeling decision here would not depend on knowing how traits are causally related among themselves, but only on the much easier to accept assumption that a trait is heritable.

For the scenario given in Figure 7, the variable *X* lies in a back-door path connecting *G* and *y*, meaning that $\mathbf{Z} = \{X\}$ fulfills the criterion while $\mathbf{Z} = \{\ \}$ does not (i.e. model c should be preferred). The back-door path is fully explained in the scenario description. Such prior knowledge may be hard (but not impossible) to achieve in some circumstances, but the decision on including or not a covariate to block a backdoor path may have simpler basis. If it is assumed that *X* may affect *y*, and if it is known that *X* and *G* are associated, but neither *G* affects *X*, nor *X* affects *G*, then the only alternative to explain such association is a back-door path (Reichenbach 1956). Accordingly, this back-door path between *G* and *y* would be assumed and represented by $G \leftrightarrow X \rightarrow y$. Under this causal assumption, *X* should be included in the model to meet the genetic effect identifiability criterion, and an explicit description of the back-door path would not be necessary to support this decision.

It is clear that this type of model choice depend on prior causal knowledge. However, relying on prior causal assumptions as a rule is required by causal inference, and given that inferring a genetic effect is learning a causal effect, one cannot avoid that issue in the inference of breeding values. It should be stressed that in many studies ignoring the identifiability issue, a model proposed could be unwittingly in accordance with the back-door criterion if it was applied. Nonetheless, to defend more formally that this model provides a sound representation of the genetic effect (which is required for selection purposes), it is necessary to go through the identifiability criterion even in the simpler settings (e.g. a scenario containing just genotypes and one phenotypic trait), since no causal meaning can be attributed to an inferred quantity if detached from causal assumptions.

In general, translating the causal knowledge to graphs makes it easier to acknowledge the causal assumptions necessary for a causal inference. The assumption of the absence of confounding due to a back-door path is probably the simpler and clearer example. Others are not so evident, because they do not involve covariate choice, but features of the data recording. For example, if the structure given in Figure 5a holds, it is true that the back-door criterion rejects the model that includes $y_2$ as a covariate. But this causal assumption also involves, for example, accepting that the data used to fit the model was not selected on the basis of $y_2$. That would not be the case if individuals with larger or smaller values of $y_2$ tended not to be included in the data set. Selecting is equivalent to conditioning, and in this case it would be the same as conditioning on a collider, activating a non-causal path and confounding the inference. The same issue applies to data that are pre-corrected for the "effects" of other variables.

**Discussion**

Improving the performance of economically important traits through selection relies on a causal relationship between genotype and phenotypes. Here, we have attempted to demonstrate that obtaining genomic predictors from fitting a genomic selection model explores an association between these two variables, but these predictors are only useful for selection if the association explored reflects a causal relationship. As the features involved in this relationship goes beyond statistics, there are no purely statistical criteria for declaring that the predictors obtained on the basis of one model are better in revealing the causal effects than predictors obtained from other models (Pearl 2000). This interpretation of genomic predictors is only possible if accompanied by assumed causal relationship among the studied variables and if this causal assumption indicates that the genetic causal effects are reflected on predictors obtained from that model. The back-door criterion is here suggested to verify that.

A considerable proportion of the efforts on genomic selection research consist in developing new models, methods and techniques in the context of animal and plant breeding. For example, many parametric and non-parametric models, as well as machine learning methods to fit the genomic signal have been proposed and compared (a comprehensive list of methods and comparisons is given by de los Campos *et al.* (2013a)). Some proposed improvements are using massive genotype data through the so-called Next-Generation Sequencing (Mardis 2008; Shendure and Ji 2008), or alternatively developing low density and cheaper SNP chips (e.g. Weigel *et al.* (2009)), possibly enriched by imputation methods (Weigel *et al.* 2010; Berry and Kearney 2011). As a general rule, the criterion to judge the quality of all methodological novelties is the genomic predictive ability. Here we try to remark that for the context of selection programs, the ability to predict in itself is not the point, as it may not be important if the signal explored does not reflect causal effects. Since the quantity to be predicted has a causal meaning, it is necessary to check first if conditions and assumptions are sufficient to infer the desired effect (not the association). Only after the genetic signal is deemed as causal, increasing the ability to predict such signal is meaningful. The very inclusion of variables with the purpose of decreasing residual variance and increasing prediction precision only matters if the causal identifiability issue is solved (Shpitser *et al.* 2012). If the goal is to infer an effect but the genomic signal explored is not causal, using alternative methodologies to increase the efficacy of separating signal from the noise does not matter for selection.

In the present study, the effects considered were additive. Therefore, the issue is not the inability to fit some complex feature of the genomic signal. One example of this other issue is using an additive model when there are dominance and/or epistasis "effects". Notice that even the proposal of models that can accommodate non-additive signal would be useful for breeding programs point of view only if the studied signal was assumed as causal. To illustrate this distinction, suppose the context of Figure 3b with dominance effects. In this case, the genetic signal from $G$ to $y$ is not additive, but under random mating the signal between $G$ and $y'$ is still expected to be additive, and this last association is interpreted as representing half of the breeding value of $G$ (Falconer 1989). Since additive effects are more relevant for selection, one may interpret that using a non-parametric predictive methodology that accommodates non-additive signal to explore the association between $G$ and $y$ may not be convenient for selection (although it could be for phenotype prediction). Nevertheless, the same method could be used for selection if the association between $G$ and $y'$ was explored. The reasoning here is that the association between $G$ and $y$ contain features that are not relevant for selection, while these features are not in the association between $G$ and $y'$, so that the genomic prediction of the offspring performance is the target information. However, this only holds if the signal between $G$ and $y'$ explored is assumed as causal. One counter-example would be fitting a model to predict the offspring performance $y_1'$ for a scenario assumed as Figure 9 using a model that considers $y_2'$ as a covariate. There is no genetic effects on $y_1'$, but including $y_2'$ in the model activates the path between the sire's genotype ($G$) and the offspring's phenotype $y_1'$. This generates a signal that could be used to predict the offspring performance from parents' genotypes. Nevertheless, this signal is not causal, and therefore is not relevant for selection. Notice that separating additive from dominance effects for selection purposes is more related to the "shape" of the signal modeled. On the other hand, in the example given based on Figure 9, notice that the lack of relevance of the genomic predictors is not a matter of signal's "shape". This last issue is the focus of this manuscript.

From the point of view of interpretation of analysis, notice that treating genomic predictions as a regression problem does not change only the meaning of genomic predictors, but also changes the meaning of other parameters. For example, following a purely predictive point of view, the estimators for the parameters traditionally named as "genetic variance" and "heritability" do not quantify variability of genetic disturbances. Such interpretation is conditional on treating predictors as reflecting genetic causal effects. If this is not the case, they could be simply seen as regularization parameters that control the flexibility of a predictive machine. This would be the case of an inferred variance parameter assigned to genomic terms from a model such as GBLUP including $y_2$ as covariate under the scenario depicted in Figure 4. This parameter would be expected to be inferred as different from 0, therefore not reflecting the heritability of that trait.

The simulated examples illustrated here were simple, and coupled with the back-door criterion they suggest that heritable traits should not be used as covariates. But this may not be the case in specific applications. Suppose an analysis of individual weaning weight ($W$) in pigs under a scenario as depicted in Figure 10. In such analysis, litter size ($LS$) could be considered as a covariate. This could be seen as a heritable trait of the individual's dam. However, the genotype of the dam ($G_m$) does not affect only the litter size, but also affect the genotype ($G$) of the individual. From that, including litter size as covariates close the back-door path between $G$ and $W$, and therefore predictors capture only the direct effect between them. But the overall graph suggests that genetics also affect weight through $LS$ (although not within a generation, and only through females), but including *LS* as covariates blocks this effect. The effect of interest (overall or direct) could be different depending on the context. Nevertheless, it is not possible to articulate this decision only on the basis of associational information (e.g. predictive ability or goodness of fit). Other example involves the inclusion of upstream traits, like in the decision involved in the scenario of Figure 6. In a standard scenario, the response to selection would depend on the overall genetic effects. But the inference of direct genetic effects would be useful when predictions are necessary for scenarios under external interventions (Valente *et al.* 2013).

Here it was shown that selection depends on a phenotypic response to an intervention on genotypes, and that the best results from this intervention depend on inferring the effect of genotypes of selection candidates on phenotypes. The response to ideally setting a genotype through intervention is the key information for selection, which using the do() notation (Pearl 2000) corresponds to $E[y|do(G)]$, which is different from the associational information $E[y|G]$. However, in selection programs, interventions on future generation's genotypes are not made through deterministic coercion to a specific value, as a parent transmits only half of its alleles, sampled from the genotype. Therefore, supposing two candidates for selection with genotypes $G_A$ and $G_B$, even if it is known with no error that the genetic effect of $G_A$ is greater, it is still possible that choosing $G_B$ as a sire results in an individual offspring with better genetic effect than choosing $G_A$. Nevertheless, selecting $G_A$ would increase the expected genetic effect, and therefore the expected phenotype, or the average phenotype of the offspring. In the sense of Eberhardt and Scheines (2007), the intervention performed in selection is not a hard intervention, but a soft one that shifts the location of a distribution. Regardless, decisions would still rely on inferring and comparing $E[y|do(G_A)]$ with $E[y|do(G_B)]$.

**References**

Akaike, H., 1973 Information theory and an extension of the maximum likelihood principle, pp. in *2nd International Symposium on Information Theory*, edited by B. N. Petrov and F. Csaki. Publishing House of the Hungarian Academy of Sciences, Budapest.

Berry, D. P., and J. F. Kearney, 2011 Imputation of genotypes from low- to high-density genotyping platforms and implications for genomic selection. Animal 5**:** 1162-1169.

de los Campos, G., D. Gianola and G. J. M. Rosa, 2009a Reproducing kernel Hilbert spaces regression: A general framework for genetic evaluation. J. Anim. Sci. 87**:** 1883-1887.

de los Campos, G., J. M. Hickey, R. Pong-Wong, H. D. Daetwyler and M. P. L. Calus, 2013a Whole-Genome Regression and Prediction Methods Applied to Plant and Animal Breeding. Genetics 193**:** 327-+.

de los Campos, G., H. Naya, D. Gianola, J. Crossa, A. Legarra *et al.*, 2009b Predicting Quantitative Traits With Regression Models for Dense Molecular Markers and Pedigree. Genetics 182**:** 375-385.

de Los Campos, G., P. Perez, A. I. Vazquez and J. Crossa, 2013b Genome-enabled prediction using the BLR (Bayesian Linear Regression) R-package. Methods in molecular biology (Clifton, N.J.) 1019**:** 299-320.

Eberhardt, F., and R. Scheines, 2007 Interventions and causal inference. Philosophy of Science 74**:** 981-995.

Falconer, D. S., 1989 *Introduction to quantitative genetics*. Longman, New York.

Fisher, R. A., 1918 The correlation between relatives on the supposition of Mendelian inheritance. Transactions of the Royal Society of Edinburgh 52**:** 399-433.

Fisher, R. A., 1971 *The design of experiments*. Macmillan, New York.

Gianola, D., 2013 Priors in Whole-Genome Regression: The Bayesian Alphabet Returns. Genetics 194**:** 573-596.

Gianola, D., and D. Sorensen, 2004 Quantitative genetic models for describing simultaneous and recursive relationships between phenotypes. Genetics 167**:** 1407-1424.

Habier, D., R. L. Fernando, K. Kizilkaya and D. J. Garrick, 2011 Extension of the bayesian alphabet for genomic selection. Bmc Bioinformatics 12.

Henderson, C. R., 1975 Best linear unbiased estimation and prediction under a selection model. Biometrics 31**:** 423-447.

Legarra, A., I. Aguilar and I. Misztal, 2009 A relationship matrix including full pedigree and genomic information. J. Dairy Sci. 92**:** 4656-4663.

Long, N., D. Gianola, G. J. M. Rosa and K. A. Weigel, 2011 Long-term impacts of genome-enabled selection. J. Appl. Genetics 52**:** 467-480.

Lynch, M., and B. Walsh, 1998 *Genetics and analysis of quantitative traits*. Sinauer, Sunderland, Mass.

Mardis, E. R., 2008 Next-generation DNA sequencing methods, pp. 387-402 in *Annual Review of Genomics and Human Genetics*.

Meuwissen, T. H. E., B. J. Hayes and M. E. Goddard, 2001 Prediction of total genetic value using genome-wide dense marker maps. Genetics 157**:** 1819-1829.

Park, T., and G. Casella, 2008 The Bayesian Lasso. J Am Stat Assoc 103**:** 681-686.

Pearl, J., 1988 *Probabilistic reasoning in intelligent systems : networks of plausible inference*. Morgan Kaufmann Publishers, San Mateo, Calif.

Pearl, J., 1995 Causal diagrams for empirical research. Biometrika 82**:** 669-688.

Pearl, J., 2000 *Causality: Models, Reasoning and Inference*. Cambridge University Press, Cambridge, UK.

Pearl, J., 2003 Statistics and causal inference: A review. Test 12**:** 281-318.

R Development Core Team, 2009 R: A Language and Environment for Statistical Computing, pp. R Foundation for Statistical Computing, Vienna, Austria.

Reichenbach, H., 1956 *The direction of time*. University of California Press, Berkeley,.

Rosa, G. J. M., and B. D. Valente, 2013 BREEDING AND GENETICS SYMPOSIUM: Inferring causal effects from observational data in livestock. J. Anim. Sci. 91**:** 553-564.

Rosa, G. J. M., B. D. Valente, G. d. l. Campos, X. L. Wu, D. Gianola *et al.*, 2011 Inferring causal phenotype networks using structural equation models. Genet. Sel. Evol. 43**:** (10 February 2011).

Schwarz, G., 1978 Estimating dimension of a model Ann. Stat. 6**:** 461-464.

Shendure, J., and H. L. Ji, 2008 Next-generation DNA sequencing. Nature Biotechnology 26**:** 1135-1145.

Shpitser, I., T. J. VanderWeele and J. M. Robins, 2012 On the validity of covariate adjustment for estimating causal effects, pp.  in *26th Conference on Uncertainty and Artificial Intelligence*. AUAI Press, Corvallis, WA.

Spiegelhalter, D. J., N. G. Best, B. R. Carlin and A. van der Linde, 2002 Bayesian measures of model complexity and fit. J. R. Stat. Soc. Ser. B-Stat. Methodol. 64**:** 583-616.

Spirtes, P., C. Glymour and R. Scheines, 2000 *Causation, Prediction and Search*. MIT Press, Cambridge, MA.

Valente, B. D., G. J. M. Rosa, D. Gianola, X. L. Wu and K. Weigel, 2013 Is Structural Equation Modeling Advantageous for the Genetic Improvement of Multiple Traits? Genetics 194**:** 561-572.

VanRaden, P. M., 2008 Efficient methods to compute genomic predictions. J. Dairy Sci. 91**:** 4414-4423.

Weigel, K. A., G. de los Campos, O. Gonzalez-Recio, H. Naya, X. L. Wu *et al.*, 2009 Predictive ability of direct genomic values for lifetime net merit of Holstein sires using selected subsets of single nucleotide polymorphism markers. J. Dairy Sci. 92**:** 5248-5257.

Weigel, K. A., G. de los Campos, A. I. Vazquez, G. J. M. Rosa, D. Gianola *et al.*, 2010 Accuracy of direct genomic values derived from imputed single nucleotide polymorphism genotypes in Jersey cattle. J. Dairy Sci. 93**:** 5423-5435.

Wu, X. L., B. Heringstad and D. Gianola, 2010 Bayesian structural equation models for inferring relationships between phenotypes: a review of methodology, identifiability, and applications. J. Anim. Breed. Genet. 127**:** 3-15.

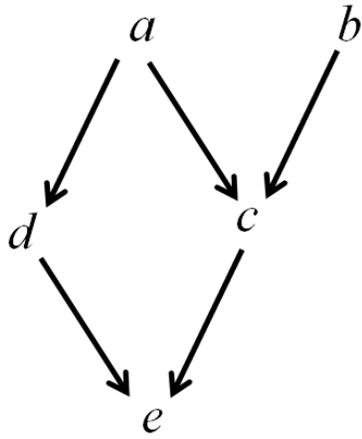

Figure 1 – A directed acyclic graph.

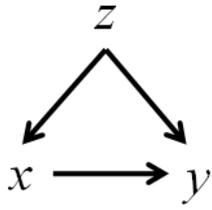 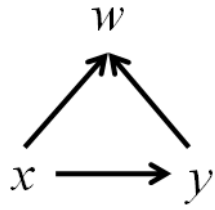

Figure 2 – Directed acyclic graphs representing hypothetical assumptions for the causal relationship between variables *x* and *y*. Nodes represent variables and directed edges represent causal effects.

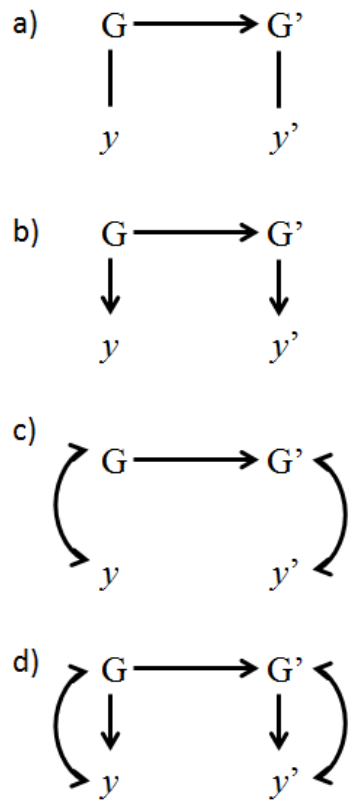

Figure 3 – Causal structures representing the selection context. The nodes *G* and *G'* represent genotypes, and the latter is assigned to a descendant of the former; *y* and *y'* are phenotypes assigned to each individual; arrows represent causal effect, bidirected arrows represent a backdoor path and undirected edges represent unresolved causal relationships.

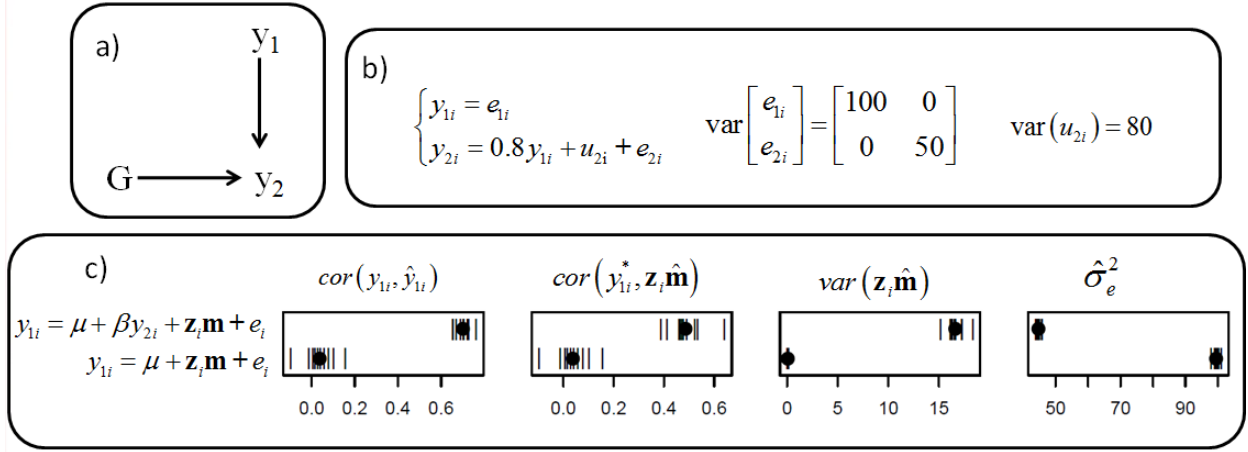

Figure 4 – Causal structure (a), causal model used for simulation (b) and results from fitting alternative models (c). In a), G represents the whole-genome genotype, and $y_1$ and $y_2$ represents two phenotypic traits. In b), $y_{1i}$ and $y_{2i}$ are phenotypes of two traits, $u_{2i}$ is the genetic effect for trait 2, $e_{1i}$ and $e_{2i}$ are residuals for respective traits, all of them assigned to the $i^{th}$ individual. In c), results are presented for predictive ability of phenotypes ($cor(y_{1i}, \hat{y}_{1i})$), and of deviations from fixed effects ($cor(y_{1i}^*, \mathbf{z}_i\hat{\mathbf{m}})$), of variability of genomic predictors ($var(\mathbf{z}_i\hat{\mathbf{m}})$) and residual variance posterior mean ($\hat{\sigma}_e^2$). Each of these results are given from fitting models ignoring ($y_{1i} = \mu + \mathbf{z}_i\mathbf{m} + e_i$) or accounting for ($y_{1i} = \mu + \beta y_{2i} + \mathbf{z}_i\mathbf{m} + e_i$) $y_{2i}$ as a covariate.

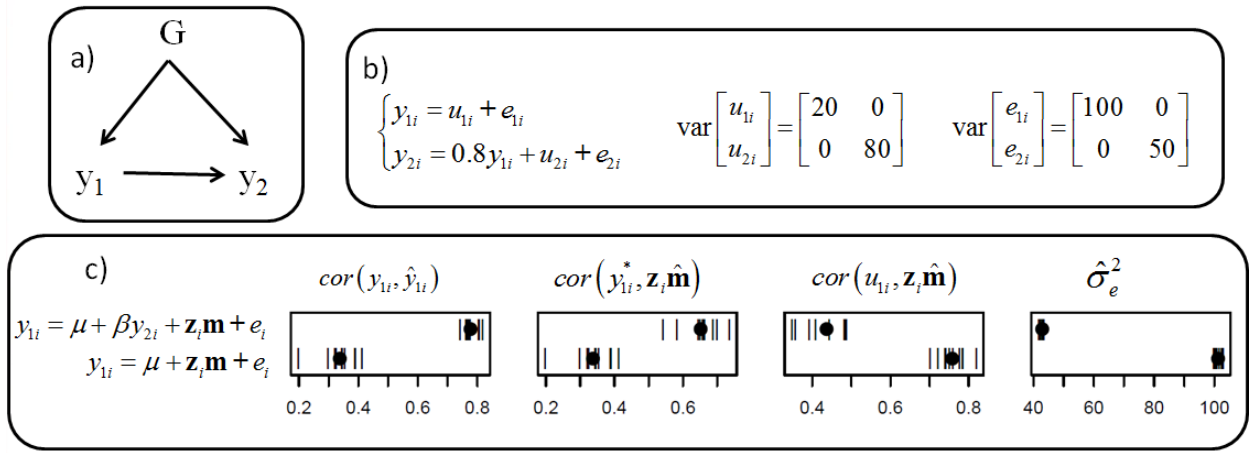

Figure 5 – Causal structure (a), causal model used for simulation (b) and results from fitting alternative models (c). In a), G represents the whole-genome genotype, and $y_1$ and $y_2$ represents two phenotypic traits. In b), $y_{1i}$ and $y_{2i}$ are phenotypes of two traits; $u_{1i}$, $u_{2i}$, $e_{1i}$ and $e_{2i}$ are genetic effects and residuals for respective traits, each one assigned to the $i^{th}$ individual. In c), results are presented for predictive ability of phenotypes ($cor(y_{1i}, \hat{y}_{1i})$), of deviations from fixed effects ($cor(y^*_{1i}, \mathbf{z}_i \hat{\mathbf{m}})$), and of the true genetic effects ($cor(u_{1i}, \mathbf{z}_i \hat{\mathbf{m}})$), as well as the residual variance posterior mean ($\hat{\sigma}^2_e$). Each of these results are given from fitting models ignoring ($y_{1i} = \mu + \mathbf{z}_i \mathbf{m} + e_i$) or accounting for ($y_{1i} = \mu + \beta y_{2i} + \mathbf{z}_i \mathbf{m} + e_i$) $y_{2i}$ as a covariate.

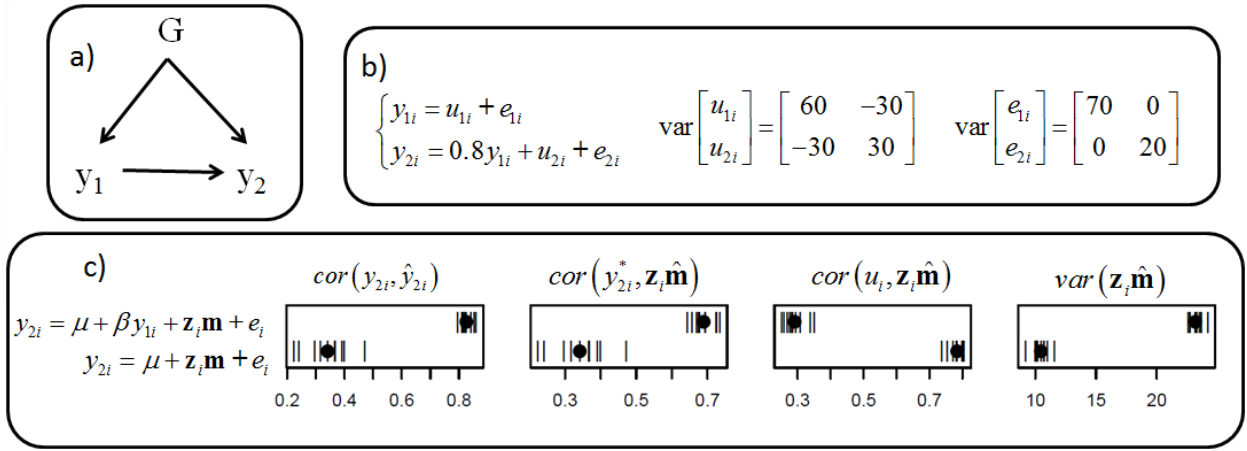

Figure 6 – Causal structure (a), causal model used for simulation (b) and results from fitting alternative models (c). In a), G represents the whole-genome genotype, and $y_1$ and $y_2$ represents two phenotypic traits. In b), $y_{1i}$ and $y_{2i}$ are phenotypes of two traits; $u_{1i}$, $u_{2i}$, $e_{1i}$ and $e_{2i}$ are genetic effects and residuals for respective traits, each one assigned to the $i^{th}$ individual. In c), results are presented for predictive ability of phenotypes ($cor(y_{2i}, \hat{y}_{2i})$), of deviations from fixed effects ($cor(y^*_{2i}, \mathbf{z}_i\hat{\mathbf{m}})$), and of the true overall genetic effects ($cor(u_i, \mathbf{z}_i\hat{\mathbf{m}})$), as well as variability of genomic predictors ($var(\mathbf{z}_i\hat{\mathbf{m}})$). Each of these results are given from fitting models ignoring ($y_{2i} = \mu + \mathbf{z}_i\mathbf{m} + e_i$) or accounting for ($y_{2i} = \mu + \beta y_{1i} + \mathbf{z}_i\mathbf{m} + e_i$) $y_{1i}$ as a covariate.

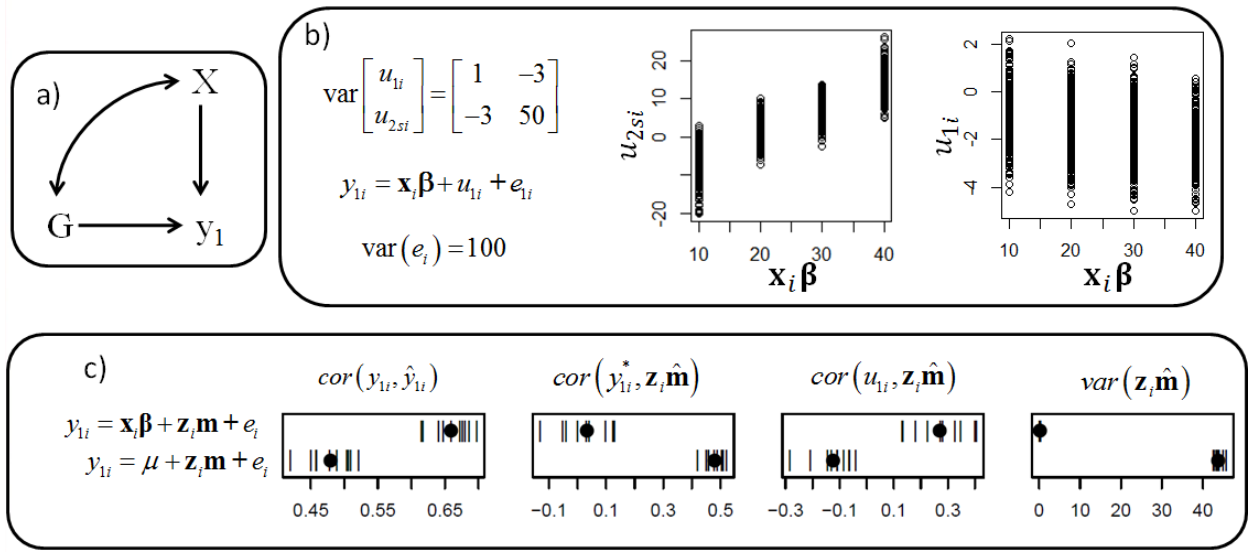

Figure 7 – Causal structure (a), causal model used for simulation (b) and results from fitting alternative models (c). In a), G represents the whole-genome genotype, $y$ represents a phenotypic trait, and $X$ represent a categorical environmental variable. The bidirected edge between $X$ and $G$ represent a back-door path. In b), $y_{1i}$, $u_{1i}$, and $e_{1i}$, are, respectively, the phenotype, genetic effect and residuals for trait 1, each one assigned to the $i^{th}$ individual; $\mathbf{x}_i\boldsymbol{\beta}$ is the fixed effects term containing categorical environmental effects on the same individual, and $u_{2si}$ is the genetic effect of the sire of the $i^{th}$ individual for trait 2. Two graphs present the dispersion of genetic effects of sires of $i$ for trait 2 and genetic effects of the $i^{th}$ individual for trait 1 for each category of $X$. In c), results are presented for predictive ability of phenotypes ($cor(y_{1i}, \hat{y}_{1i})$), of deviations from fixed effects ($cor(y_{1i}^*, \mathbf{z}_i\hat{\mathbf{m}})$), and of the true genetic effects ($cor(u_{1i}, \mathbf{z}_i\hat{\mathbf{m}})$), as well as variability of genomic predictors ($var(\mathbf{z}_i\hat{\mathbf{m}})$). Each of these results are given from fitting models ignoring ($y_{1i} = \mu + \mathbf{z}_i\mathbf{m} + e_i$) or accounting for ($y_i = \mathbf{x}_i\boldsymbol{\beta} + \mathbf{z}_i\mathbf{m} + e_i$) $X$ as a covariate.

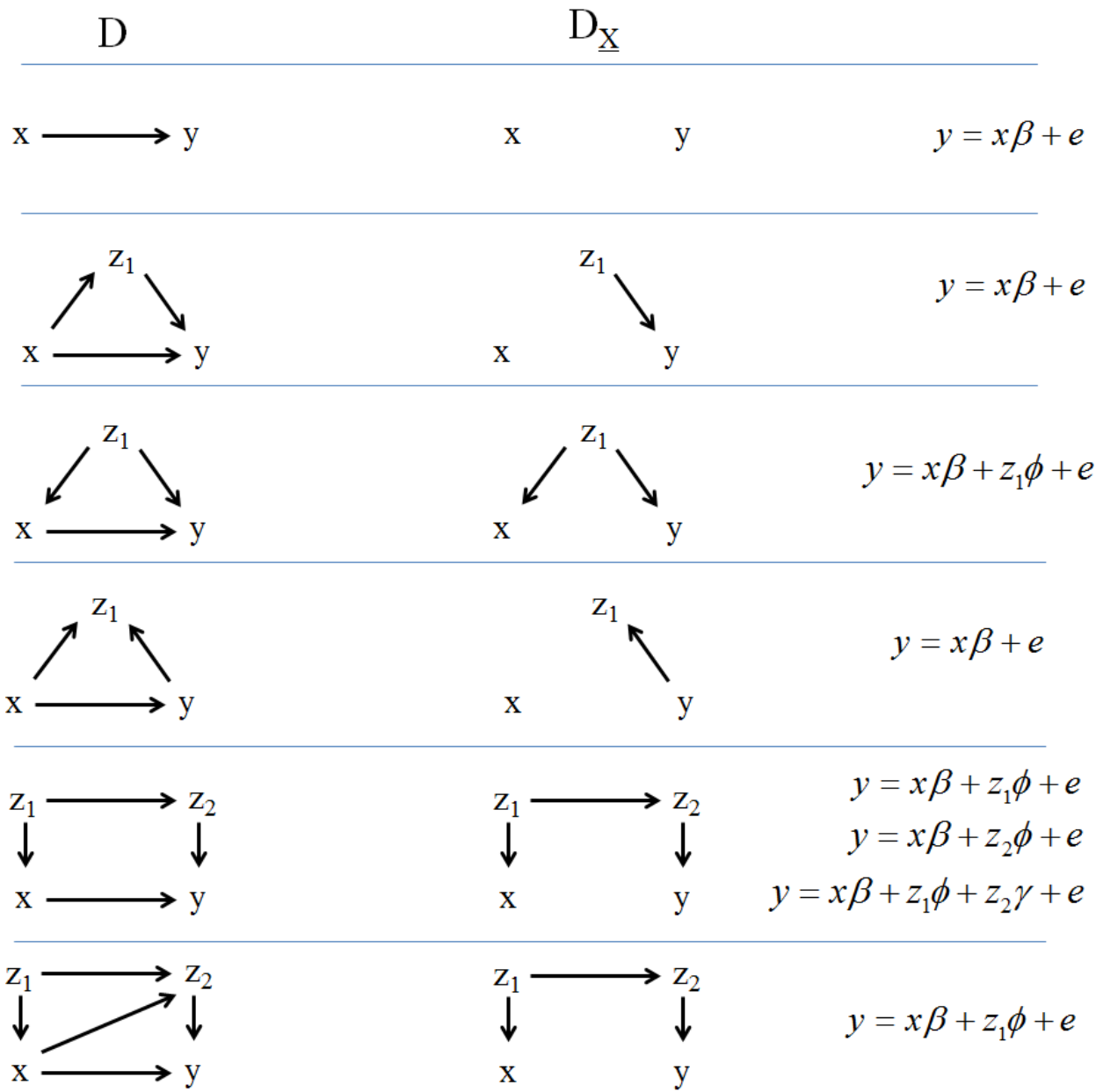

Figure 8 – Examples of the application of the back-door criterion to covariate selection when the goal is estimating the effect of *x* on *y*. Graphs in **D** are the assumed causal relationship between *x* and *y*, **D**$_{\underline{x}}$ is **D** devoid of arrows emanating from *x*. The models listed in the right side are constructed following the back-door criterion, so that $\hat{\beta}$ can be treated as an estimated effect.

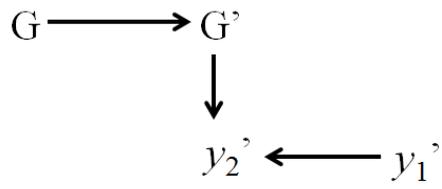

Figure 9 – Causal structure representing relationships between genotypes from an individual and its offspring ($G$ and $G'$) and two traits expressed in the offspring ($y_1'$ and $y_2'$); Arrows represent causal effects.

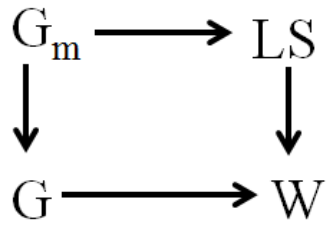

Figure 10 – Causal structure representing relationships among weaning weight (*W*), litter size (*LS*), the genotype of an individual (*G*) and of its dam ($G_m$). Arrows represent causal connections.